\documentclass{revtex4}

\usepackage{cmap}
\usepackage[latin1]{inputenc}
\usepackage[T1]{fontenc}
\usepackage{microtype,booktabs}
\usepackage[english]{babel}
\usepackage{graphicx}

\usepackage{amsmath,amssymb,amsfonts,amsthm}
\usepackage{multirow}

\usepackage{hyperref}

\newcommand{\prob}[2][]{ \mathbb{P}_{#1}\left(#2\right)}
\newcommand{\probhat}[2][]{ \widehat{\mathbb{P}}_{#1}\left(#2\right)}

\newtheorem{theorem}{Theorem}[section]
\newtheorem{assumption}[theorem]{Assumption}

\begin{document}
 
\title{Bethe Ansatz for the Weakly Asymmetric Simple Exclusion Process and phase transition in the current distribution}

\author{Damien Simon}
\email{damien.simon@upmc.fr}
\affiliation{Laboratoire Probabilit\'es et mod\`eles al\'eatoires, UMR 7599, UPMC Paris 6 et CNRS, 4 place Jussieu, 75252 Paris Cedex 05}

\begin{abstract}
The probability distribution of the current in the asymmetric simple exclusion process is expected to undergo a phase transition in the regime of weak asymmetry of the jumping rates. This transition was first predicted by Bodineau and Derrida using a linear stability analysis of the hydrodynamical limit of the process and further arguments have been given by Mallick and Prolhac. However it has been impossible so far 
to study what happens after the transition. The present paper presents an analysis of the large deviation function of the current on both sides of the transition from a Bethe Ansatz approach of the weak asymmetry regime of the exclusion process. 
\end{abstract}

\maketitle

\section{Introduction}
The simple exclusion process is one of the toy-models of out-of-equilibrium statistical physics and was introduced first around forty years ago \cite{spitzer}. Many aspects of equilibrium statistical physics of particle systems are now well understood, however only a few approaches are known for out-of-equilibrium systems. Such systems usually have a non-trivial irreversible dynamics that prevents the stationary regime to be Boltzmann-Gibbs-like. Moreover, they are also characterized by the existence of non-zero currents (energy, mass, etc.) going through the system. 

The asymmetric simple exclusion process (ASEP), as defined below, is such a model of mass transport along a one-dimensional periodic track of length $L$: each site is occupied by at most one particle and each particle jumps to the right (resp. left) with a rate $p$ (resp. $q$) if the target site is empty. The exclusion rule induces an interaction between neighbouring particles. If $p\neq q$, a current of particles flows through the system. 

Close to equilibrium, for example if the difference $p-q\propto 1/L$ is small on a large lattice, the system can reach \emph{local} equilibrium and it can be described at the macroscopic scale by a hydrodynamical theory \cite{bertini-et-al1,bertini-et-al2,bertini-et-al3,bertini-et-al4,bodineau-additivity}. This theory allows in principle to compute the typical macroscopic density profile that corresponds to the observation of a given current as well as the large deviation of this current from a variational principle. Finding the global optimal profile in this framework remains however a difficult task \cite{bodineau-derrida,bodineau-derrida2}: the flat density profile is the optimal profile for currents close to the mean current but a linear stability analysis shows that this optimal profile becomes unstable and time-dependent profile for smaller currents. Although the scenario of a traffic jam occurring at the transition seems clear, the characterization of the transition and of the jammed profile was still unknown.

\begin{figure}
\includegraphics[width=5cm]{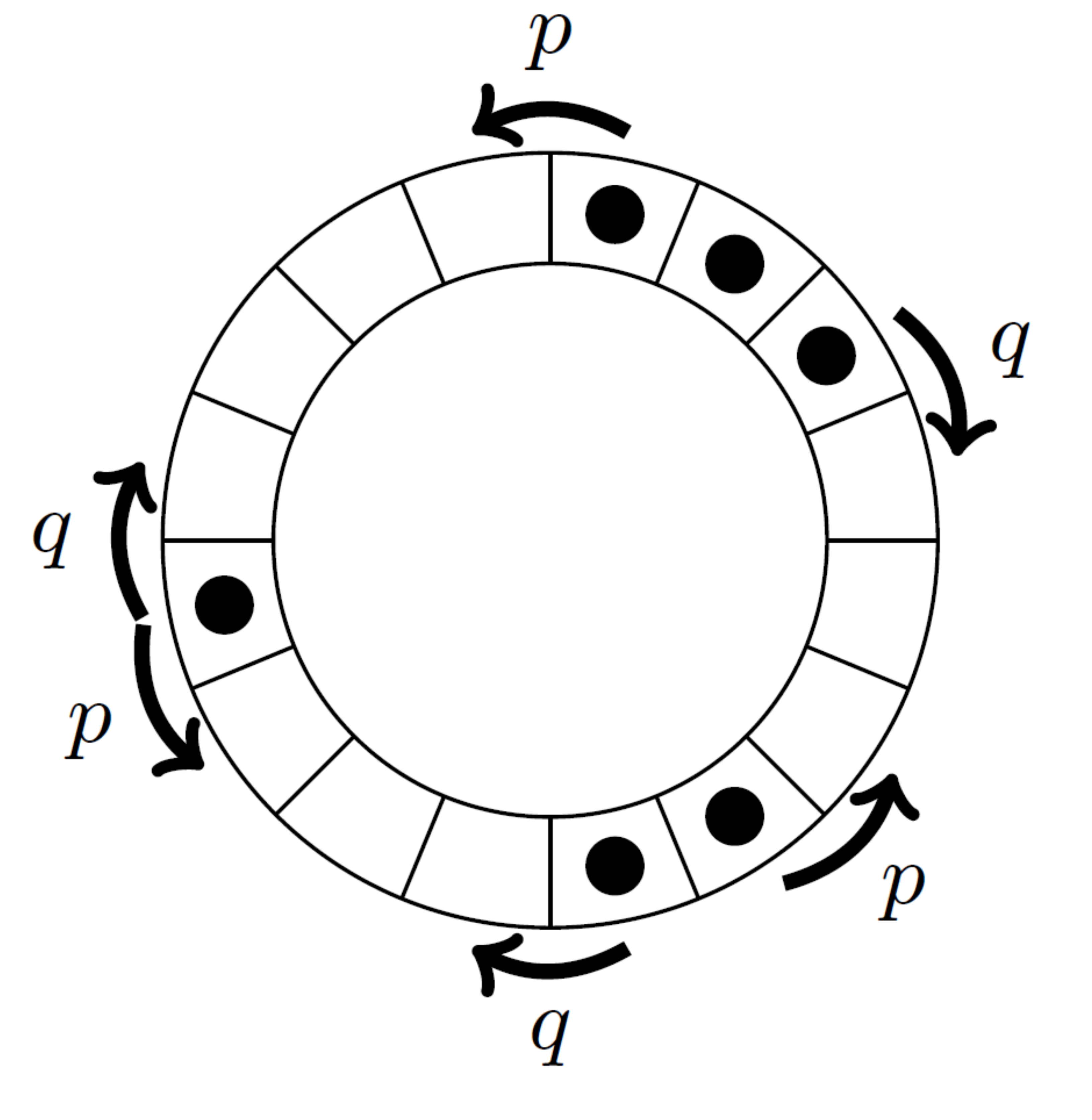}
\caption{\label{fig:asepschema}Asymmetric exclusion process on a ring geometry of $L=16$ sites with $N=6$ particles}
\end{figure}

A second approach to the asymmetric exclusion process comes from the integrability of the microscopic model and the relation between its Markov matrix and the Hamiltonian of the XXZ quantum spin chain. Formally, the Markov matrix modified so that the current can be counted can be diagonalized with the Bethe Ansatz and the eigenvalues parametrized by complex numbers satisfying coupled non-linear equations. This approach has already lead to many results about the stationary state of exclusion process  and the slowest relaxation times on the ring \cite{gwaspohn1,gwaspohn2,kim1,kim2,dorlas-priezzhev,Priezzhev1} and more recently with couplings to two reservoirs \cite{degieressler1,degieressler2,degieressler3,simon}. It has been also possible to study the fluctuations of the current both numerically \cite{kurchanpeliti,tailleurlecomte} and theoretically through the systematic perturbative computation of the cumulants of the current \cite{derrida-vanw-lecomte,prolhac-cumulants,prolhac-cumul-weak,prolhac-tree,prolhac-these}. The present paper shows how to go beyond these results and describe the jammed phase as well as the transition itself from the point of view of the Bethe Ansatz.

The paper is organized as follows: section \ref{sec:generalprops} introduces the model, the notations and some previous results. In section \ref{sec:betheeqs}, the Bethe equations are rewritten in a suitable way to take the simultaneous limit of large size and weak asymmetry. The Gaussian part of the probability distribution of the current is recovered in section \ref{sec:gaussianpart} from the Bethe Ansatz, the transition in section \ref{sec:phasetransition} and the jammed phase in \ref{sec:jammedphase}, followed by a more general discussion of the type of phase transition that is described.

\section{General properties}
\label{sec:generalprops}
	\subsection{Definitions}
The asymmetric exclusion process describes $N$ particles evolving on a one-dimensional lattice of length $L$ with periodic boundary conditions (see figure \ref{fig:asepschema}). Each site $i$ is occupied by at most $1$ particle and is described by an occupation number $\tau_i\in\{0,1\}$. The positions of the particles are noted $x_j$ for $1\leq j \leq N$ so that $\tau_i=1$ if and only if $i=x_j$ for some $j$. 

During a infinitesimal time $dt$, a particle on a site $i$ tries to jump with rate $p$ to the site $i+1$ and with rate $q$ to the site $i-1$: if the target site is empty, the jump occurs, whereas it is cancelled if it is occupied. In the two-site basis $(00,01,10,11)$, the Markov matrix that describes the jumps through the corresponding edge is
\begin{equation}
w=
\begin{pmatrix}
0 & 0 & 0 &0 \\
0 & -q & p & 0 \\
0 & q & -p & 0 \\
0 & 0 & 0 & 0
\end{pmatrix}.
\end{equation} 
The Markov stochastic matrix $\mathcal{W}$ for the whole lattice is then given by
\begin{equation}
\mathcal{W}= \sum_{i=1}^L w_{i,i+1},
\end{equation}
where the site $L+1$ is identified to the site $1$ and $w_{i,i+1}$ is the matrix $w$ above acting on the pair of sites $(i,i+1)$. This dynamics keeps the number of particles constant and one can label each sector of $W$ by the density $\rho_0=N/L$.

A slight modification of the matrix $\mathcal{W}$ gives the Legendre transform of the large deviation function of the current. The current $Q$ accumulated between $0$ and $t$ is defined as the difference between the total numbers of jumps of particles to the right and of jumps to the left. The probability $\prob[t]{\mathcal{C},Q}$ of observing the configuration $\mathcal{C}$ at time $t$ and a total current $Q$ between $0$ and $t$ satisfies
\begin{equation}
\label{eq:mastereq}
\partial_t \prob[t]{\mathcal{C},Q} = \sum_{\mathcal{C}'\neq \mathcal{C}} \mathcal{W}_{\mathcal{C}\mathcal{C'}} \prob[t]{\mathcal{C}',Q-q_{\mathcal{C'},\mathcal{C}}} - \left( \sum_{\mathcal{C}'\neq \mathcal{C}} \mathcal{W}_{\mathcal{C'}\mathcal{C}}\right) \prob[t]{\mathcal{C},Q},
\end{equation}
where $\mathcal{W}_{\mathcal{C}\mathcal{C'}}$ is the matrix element of $\mathcal{W}$ which gives the transition rate from $\mathcal{C'}$ to $\mathcal{C}$ and $q_{\mathcal{C'},\mathcal{C}}\in\{-1,1\}$ is the increment of the current corresponding to the change from $\mathcal{C'}$ to $\mathcal{C}$. The generating function of the current $\probhat[t]{\mathcal{C},s}$ defined as
\begin{equation}
\probhat[t]{\mathcal{C},s} = \sum_{Q\in \mathbb{Z}} e^{sQ} \prob[t]{\mathcal{C},Q}
\end{equation}
satisfies a modified version of \eqref{eq:mastereq}:
\begin{equation}
\partial_t \probhat[t]{\mathcal{C},s} = \sum_{\mathcal{C}'\neq \mathcal{C}} \Big( e^{sq_{\mathcal{C},\mathcal{C}'}} \mathcal{W}_{\mathcal{C}\mathcal{C'}}\Big) \probhat[t]{\mathcal{C}',s} - \left( \sum_{\mathcal{C}'\neq \mathcal{C}} \mathcal{W}_{\mathcal{C'}\mathcal{C}}\right) \probhat[t]{\mathcal{C},s}.
\end{equation}
This equation is linear and its linear operator $\mathcal{W}^{(s)}$ is given by
\begin{eqnarray}
\mathcal{W}^{(s)} &=& \sum_{i=1}^{L+1} w_{i,i+1}^{(s)}, \\
w_{i,i+1}^{(s)} &=&
\begin{pmatrix}
0 & 0 & 0 & 0 \\
0 & -q & pe^s & 0 \\
0 & qe^{-s} & -p & 0 \\
0 & 0 & 0 & 0
\end{pmatrix}.
\end{eqnarray}
The matrix $\mathcal{W}^{(s)}$ is not stochastic any longer but its eigenvalue $\lambda_1(s)$ with the maximal real part is real and non-degenerate (Perron-Frobenius' theorem) in each sector $N$. This eigenvalue gives the long time behaviour $\probhat[t]{\mathcal{C},s}\propto \phi_1^{(s)}(\mathcal{C}) e^{\lambda_1(s) t}$ where $\phi_1^{(s)}(\mathcal{C})$ is the associated eigenvector. This exponential growth corresponds to a large deviation behaviour of the distribution of the current:
\begin{equation}
\frac{\ln \prob[t]{\mathcal{C},jt}}{t} \xrightarrow{t\to\infty} f(j)
\end{equation}
where $f(j)$ and $\lambda_1(s)$ are related through the Legendre transform
\begin{equation}
\label{eq:legendre}
\lambda_1(s) = \sup_{j} \Big( f(j)+sj\Big).
\end{equation}
The expected phase transition described in the introduction \cite{bodineau-derrida,bodineau-derrida2} occurs in the function $f(j)$ (resp. $\lambda_1(s)$) as the parameter $j$ (resp. $s$) varies. The function $f(j)$ is negative and is zero when $j$ is equal to the mean current. The successive derivatives of $\lambda_1(s)$ at $s=0$ give the cumulants of the current $\langle Q_t^n\rangle_c$ already given for some cases in \cite{prolhac-cumulants,prolhac-cumul-weak,prolhac-these}. 

Once normalized, the eigenvector $\phi_1^{(s)}(\mathcal{C})$ describes the stationary measure of the configurations conditioned on the \emph{past} current. 
The matrix $\mathcal{W}^{(s)}$ has left eigenvectors $\psi_i^{(s)}(\mathcal{C})$ associated to the same eigenvalues as the right eigenvectors $\phi_i^{(s)}(\mathcal{C})$. For $s=0$, the stochasticity of $\mathcal{W}$ implies that the eigenvector associated to $\lambda_1(0)=0$ is given by $\psi_1^{(0)}(\mathcal{C})=1$ up to a normalization constant. As explained for example in \cite{simon}, the knowledge of $\psi_1^{(s)}(\mathcal{C})$ is necessary to understand the distribution of the configurations conditioned to the production of a given current $j$ at \emph{larger} times. The probability  of observing a configuration $\mathcal{C}$ at time $t= \alpha T$ for some $\alpha $ given in $]0,1[$ knowing that the current at the final time $T$ is equal to $Q_T=jT$ converges as $T\to\infty$ to $\mathbb{P}_\text{cond}^{(j)}(\mathcal{C})$ given by:
\begin{equation}
\mathbb{P}_\text{cond}^{(j)}(\mathcal{C}) = \frac{1}{Z} \psi_1^{(s_j)}(\mathcal{C}) \phi_1^{(s_j)}(\mathcal{C})
\end{equation}
where $s_j=-f'(j)$ and $Z$ is a normalization constant (see \cite{simon}).

\paragraph*{Symmetries.} The asymmetric exclusion process is invariant under the particle-hole transformation. For a system of $N$ particles, the $L-N$ holes jump to the left with rate $p$, to the right with rate $q$ and satisfy the exclusion rules. The system is also translation-invariant. A less trivial symmetry is the Gallavotti-Cohen symmetry \cite{gallavotti} that relates the probabilities of time-reversed histories of the system. One can check on the definition of $\mathcal{W}^{(s)}$ that it satisfies:
\begin{equation}
{\mathcal{W}^{(s)}}^T = \mathcal{W}^{(s')}, \quad\text{with $s+s'+\ln(p/q)=0$} 
\end{equation}
where $M^T$ denotes the transposition of the matrix $M$. It follows that the first eigenvalue $\lambda_1(s)$, which is non-degenerate, satisfies:
\begin{equation}
\lambda_1(s) = \lambda_1(-\ln(p/q)-s).
\end{equation}
This identity in terms of the large deviation function $f(j)$ can be written $f(j)-f(-j)= \ln(p/q) j$.

\paragraph*{Scalings.} The following sections give the largest eigenvalue $\lambda_1(s)$ for any $s$ in the limit of weak asymmetry defined as:
\begin{subequations}
\label{eq:wasepregime}
\begin{eqnarray}
p  &=& \frac{1}{2} + \frac{\nu}{2L} + o\left(L^{-1}\right), \\
q  &=& \frac{1}{2} - \frac{\nu}{2L} + o\left(L^{-1}\right), \\
s &=& \frac{\gamma}{L} + o(L^{-1}).
\end{eqnarray}
\end{subequations}
This scaling corresponds to the scaling already used in many different works (see \cite{bodineau-derrida,prolhac-cumul-weak} for examples) but gives very \emph{different} results from the scaling $p-q\propto 1/\sqrt{L}$, which is  often also called the weak asymmetry regime and is related to the KPZ universality class \cite{bertini-giacomin,sasamoto-spohn,quastel}.

Remark that this scaling is not usual in spin chain models such as the XXZ spin chain, for which the coupling constants are not assumed to depend on the size of the system. However, in the present case, the scaling is natural from the point of view of the discretization of a macroscopic continuous diffusion process.

	\subsection{Hydrodynamical description}
	\label{sec:hydro}
	
In the weak asymmetry regime \eqref{eq:wasepregime}, the system can be described on the macroscopic scale by the hydrodynamic limit studied in \cite{bertini-et-al1,bodineau-derrida,bodineau-derrida2}. A macroscopic configuration is characterized by its density profile $\rho(x,t)$ at time $t$ where $x\in [0,1]$ is the rescaled position. The large deviation function $f(j_0)$ corresponding to the observation of a current $j_0$ over a large time $T$ is given in this approach by a variational principle \cite{bertini-et-al1}:
\begin{equation}
\label{eq:hydro:variational}
f(j_0)   \underset{\text{$L$ large}}{\simeq} \frac{1}{L} \lim_{T \to \infty} \left( -\frac{1}{T} \inf_{\rho(x,\tau)} \mathcal{I}_{[0,T]}(j,\rho)\right)
\end{equation}
where the local current $j(x,\tau)$ is an auxiliary variable which must satisfy both the global current constraint
\begin{equation}
\label{eq:currentfixed}
\lim_{T\to\infty} \frac{1}{T}\int_0^T j(x,\tau) d\tau = j_0
\end{equation}
and the local conservation of the mass
\begin{equation}
\label{eq:massconserv}
\partial_\tau \rho(x,\tau) + \partial_x j(x,\tau) = 0.
\end{equation}
The position- and time-dependent density profile must also satisfy the conservation law $\int_0^1 \rho(x,\tau) dx = \rho$. The functional $\mathcal{I}_{[0,T]}(j,\rho)$ is given for the weakly asymmetric simple exclusion process by:
\begin{eqnarray}
\mathcal{I}_{[0,T]}(j,\rho) &=& \int_0^T d\tau \int_0^1 dx \frac{\Big( j(x,\tau)+D(\rho(x,\tau))\partial_x\rho(x,\tau) - \nu \sigma(\rho(x,\tau))\Big)^2}{2\sigma( \rho(x,\tau))}, \label{eq:actionI}\\
D(\rho) &=& \frac{1}{2} ,\\
\sigma(\rho) &=& \rho(1-\rho).
\end{eqnarray}
The two functions $D$ and $\sigma$ given in this equation are particular to the WASEP but the expression of $\mathcal{I}_{[0,T]}(j,\rho)$ is more general and can describe other non-equilibrium diffusive systems \cite{bertini-et-al1,bertini-et-al4}.

Minimizing \eqref{eq:hydro:variational} is not easy since the optimal profile may become time-dependent. For $j$ close to the mean current, the flat solution $\rho(x,\tau)=\rho_0$ is found to be the optimal profile \cite{bodineau-derrida}. In this case, a simple computation shows that the distribution of the current is nearly Gaussian and thus one has $L f(j_0) \to -(j_0-\nu\sigma(\rho_0))^2/(2\sigma(\rho_0))$ and
\begin{equation}
\label{eq:hydro:gaussianlargedev}
L \lambda_1(s=\gamma/L)  \xrightarrow{L\to\infty} \tilde{\lambda}_1^\text{gauss}(\gamma)= \frac{\sigma(\rho_0)}{2}\left(  \gamma^2+ 2\nu \gamma \right).
\end{equation}
A phase transition is however expected to occur when $j=j_c$ (and $\gamma=\gamma_c$ for $\tilde{\lambda}_1(\gamma)$) since it has been shown in \cite{bodineau-derrida} that after $j_c$ the flat profile is linearly unstable for the evolution induced by $\mathcal{I}_{[0,T]}(j,\rho)$. The critical parameter $\gamma_c$ is obtained by solving 
\begin{equation}
\label{eq:hydrotransition}
\tilde{\lambda}_1^\text{gauss}(\gamma) = -\pi^2/2.
\end{equation}

For $\tilde{\lambda}_1^\text{gauss}(\gamma)<-\pi^2$, the characterization of the optimal profile is still missing although candidates with a travelling wave shape $\rho(x,\tau)=g(x-v\tau)$ have been studied \cite{bodineau-derrida,bodineau-derrida2} and found to produce a lower value of $\mathcal{I}_{[0,T]}(j,\rho)$ than the flat profile. The minimization of \eqref{eq:actionI} over the moving profiles $g$ on the ring and the velocities $v$ leads to a profile $g$ determined only implicitly from a set of elliptic integrals. The details of the equations that are obtained are presented in appendix \ref{app:profiles}.

	\subsection{Integrability of the microscopic dynamics}
	
From an algebraic point of view, the matrix $\mathcal{W}^{(s)}$ can be mapped onto the Hamiltonian of an XXZ spin chain with periodic twisted non-hermitian boundary conditions, which is known to be integrable (see \cite{gwaspohn1,gwaspohn2,kim1,kim2,derrida-appert} for earlier results on ASEP-like models based on Bethe Ansatz methods). The only result from integrability used in this section is that the matrix $\mathcal{W}^{(s)}$ can be diagonalized with the coordinate Bethe Ansatz method~\cite{gaudin}. The eigenvector for a system of $N$ particles can be parametrized by $N$ complex numbers $z_k$ with
\begin{equation}
\label{eq:betheansatz}
\phi_i^{(s)}( \mathbf{x}_< ) = \sum_{ \sigma \in \mathcal{S}_N} A_\sigma(\{z_l\}) \prod_{k=1}^N z_{\sigma(k)}^{x_k}
\end{equation}
where the sum is performed over all the permutations $\sigma$ of $N$ elements and $\mathbf{x}_<=(x_1,\ldots,x_N)$ is the vector containing the ordered positions $1\leq x_1 < x_2 < \ldots < x_N \leq L$ of the $N$ particles on the ring. The coefficients $A_P$ and the $N$ numbers $z_k$ are fixed by imposing the eigenvalue condition $\mathcal{W}^{(s)}\phi_i^{(s)} = \lambda_i^{(s)}\phi_i^{(s)}$, up to a global normalization constant of the vector $\phi_i^{(s)}$.

The $N$ Bethe equations resulting from the Ansatz \eqref{eq:betheansatz} are given for all $j\in\{1,\ldots,N\}$ by:
\begin{equation}
\label{eq:betheeqs:z}
z_j^L = (-1)^{N-1} \prod_{k\neq j} \frac{ pe^s + qe^{-s} z_kz_j -(p+q)z_j}{ pe^s + qe^{-s} z_kz_j -(p+q)z_k}
\end{equation}
with the eigenvalue:
\begin{equation}
\label{eq:eigenvalueBethe:z}
\lambda = \sum_{k=1}^N \left( \frac{pe^s}{z_k} + qe^{-s} z_k -(p+q) \right).
\end{equation}
One can read \cite{prolhac-cumulants,simon} for easy derivations of these equations and eigenvalues, as well as for the expressions of the coefficients $A_\sigma(\{z_k\})$.

This set of non-linear equations has many solutions but is expected to contain all the eigenvalues and thus the maximal one $\lambda_1(s)$. For $s=0$, the matrix $\mathcal{W}^{(s)}$ is stochastic and the first eigenvalue is known to be $\lambda_1(0)=0$; moreover, the corresponding eigenvector has constant coordinates since all the configurations are shown to have the same probability. In terms of the Bethe roots, it corresponds to the Bethe roots $z_k$ going all to the same value $1$ as $s\to 0$. A perturbative study of $\lambda_1(s)$ around $s=0$ for all sizes $L$ done in \cite{prolhac-cumulants,prolhac-cumul-weak,prolhac-these} has already given the expression of the cumulants of the current for the ASEP for various asymmetry regimes and the first finite-size corrections. In particular, the attempt of resummation of the large deviation function $f(j)$ from the cumulants has shown that a singularity (finite convergence radius) occurs for the same critical values $j_c$ and $s_c$ as in the hydrodynamic approach. Nevertheless, the perturbative expansion does not allow to understand what happens in the jammed phase, nor at the phase transition.

The translation invariance of $\mathcal{W}^{(s)}$ implies that the eigenvectors $\phi_i^{(s)}$ are eigenstates of the translation operator. It is indeed the case as it can be seen from the construction \eqref{eq:betheansatz} and the eigenvalue is the product $z_1\ldots z_N$. One can check from \eqref{eq:betheeqs:z} that one has $(z_1\ldots z_N)^L=1$, which corresponds to the periodic boundary conditions. In the case of the first eigenvalue $\lambda_1(s)$, the eigenvector $\phi_1^{(s)}$ is non-degenerate and all its entries are positive (Perron-Frobenius); moreover it has to be an eigenvalue of the one-site translation operator with eigenvalue $\exp(2i\pi n/L)$, $n\in\{0,\ldots,L-1\}$: the reality and the positivity of the entries of $\phi_1^{(s)}$ select together the integer $n=0$ for $\lambda_1(s)$ and one has to satisfy
\begin{equation}
\label{eq:translationeigen}
\prod_{k=1}^N z_k = 1
\end{equation}
for the first eigenvalue $\lambda_1(s)$.
The Gallavotti-Cohen symmetry \cite{gallavotti} corresponds to the change of variable $e^s \mapsto (q/p)e^{-s}$ and $z_k\mapsto 1/z_k$.

\section{Bethe equations for the WASEP}
\label{sec:betheeqs}
	\subsection{Polynomial representation}
	
The form \eqref{eq:betheeqs:z} of the Bethe equations is not suitable to a simple analysis of the weak asymmetry regime $s\simeq \gamma/L$ and $p/q\simeq1+2\nu/ L$. The change of variable
\begin{equation}
z_k = e^{s} \frac{ u_k \sqrt{p/q}+ 1}{u_k \sqrt{q/p} +1 }
\end{equation}
inserted in \eqref{eq:betheeqs:z} leads to the following set of equations on the $u_k$'s:
\begin{equation}
\label{eq:betheeqs:u}
\forall j\in \{1,\ldots,N\}, \quad e^{sL} \left( \frac{u_j\sqrt{p/q}+1}{u_j\sqrt{q/p}+1}\right)^L = \prod_{k\neq j} \frac{pu_j- q u_k}{qu_j-pu_k }.
\end{equation}
The change of variable is such that $u_j$ goes to $0$ as $s\to 0$ and $z_k\to 1$ simultaneously and is defined as long as $z_k\neq e^s (p/q)$ and $u_k\notin \{-\sqrt{p/q},-\sqrt{q/p}\}$. The Gallavotti-Cohen symmetry corresponds to the change $u_k \mapsto 1/u_k$. One notices that the interaction terms in the r.h.s. of \eqref{eq:betheeqs:z} and \eqref{eq:betheeqs:u} are now rational fractions with numerator and denominator of degree $1$. The polynomial $Q_N(U)$, of degree $N$, is defined as
\begin{equation}
Q_N(U) = \prod_{k=1}^N (U-u_k),
\end{equation}
so that the set of $N$ equations \eqref{eq:betheeqs:u} is equivalent to the divisibility of the polynomial 
\begin{equation*}
e^{sL}p^N \left( U\sqrt{\frac{p}{q}}+1\right)^L Q_N\left( \frac{q}{p} U\right) + q^N \left( U\sqrt{\frac{q}{p}}+1\right)^L Q_N\left( \frac{p}{q} U\right)
\end{equation*}
by $Q_N(U)$, since they are both vanishing on the $u_j$, that are supposed to be distinct in the generic case. Thus, there exists a polynomial $R_{L,N}(U)$ of degree $L$ such that~:
\begin{equation}
\label{eq:betheeqs:polyn}
R_{L,N}(U) Q_N(U) = e^{sL}p^N \left( U\sqrt{\frac{p}{q}}+1\right)^L Q_N\left( \frac{q}{p} U\right) +q^N \left( U\sqrt{\frac{q}{p}}+1\right)^L Q_N\left( \frac{p}{q} U\right).
\end{equation}
The r.h.s. of this equation is linear in $Q_N$ but is non-local in $U$ since it implies the computation of $Q_N$ at other points than $U$. There does not exist, to our knowledge, any systematic method to solve algebraic equations such as \eqref{eq:betheeqs:polyn} for all values of $L$ and $N$. Only asymptotic solutions for $L,N\to\infty$ are available under some additional hypotheses on the asymptotics of the Bethe roots, as well as some relations with Askey-Wilson polynomials \cite{sturm-liouville}.

The translation condition \eqref{eq:translationeigen} can be written in terms of $Q_N(U)$ for the eigenvalue $\lambda_1(s)$:
\begin{equation}
\label{eq:translation:Qrep}
e^{sN} \left(\frac{p}{q}\right)^N \frac{Q_N(-\sqrt{q/p})}{Q_N(-\sqrt{p/q})} = 1.
\end{equation}	
Eq.~\eqref{eq:betheeqs:polyn} for $U=0$ gives the normalization condition
\begin{equation}
 \label{eq:R0}
 R_{L,N}(0) = e^{sL} p^N+ q^N
\end{equation}
of the polynomial $R_{L,N}$.

	\subsection{The resolvent and its properties}

The purpose of this sequence is to translate the algebraic divisibility property of $Q_N$ for finite $N$ and $L$ into analytic properties when $L,N\to\infty$ with $N/L\to\rho_0$ and the weak asymmetry scaling.
Eq.~\eqref{eq:betheeqs:polyn} can be rewritten~:
\begin{equation}
\label{eq:bethepolyn:bis}
\begin{split}
\frac{R_{L,N}(U)}{2(\sqrt{pq})^N(1+U)^L} e^{-sL/2}= & \frac{1}{2}\Bigg[e^{sL/2+N\ln \sqrt{p/q}} \left(\frac{1+\sqrt{p/q}U}{1+U}\right)^L \frac{Q_N((q/p)U)}{Q_N(U)} ,\\
&+ e^{-sL/2-N\ln\sqrt{p/q}} \left(\frac{1+\sqrt{q/p}U}{1+U}\right)^L \frac{Q_N((p/q)U)}{Q_N(U)} \Bigg] .
\end{split}
\end{equation}
The ratios $Q_N(V)/Q_N(U)$ can be rewritten with the so-called resolvent $W_{L,N}(U)$:
\begin{subequations}
\label{eq:betheeqs:newdefs}
\begin{eqnarray}
\frac{Q_N(V)}{Q_N(U)} &=& \exp\left( L\int_U^V W_{L,N}(u) du\right) ,\\
W_{L,N}(U) &=& \frac{1}{L}\sum_{k=1}^N \frac{1}{U-u_k}.
\end{eqnarray}
\end{subequations}
The function $W_{L,N}(U)$ has $N$ poles located at the $u_k$'s, is holomorphic everywhere else on $\mathbb{C}$ and decreases as $\rho_0/U$ as $|U|\to\infty$ where $N/L=\rho_0$.

The resolvent technique presented here is standard in the study of the thermodynamic limit of Bethe equations and a simple overview of this technique in a similar algebraic context is presented in \cite{gromovkazakov}.

In most systems solved by coordinate Bethe Ansatz, the Bethe roots that correspond to the ground state $\lambda_1$ condensate on a finite number of smooth curves or degenerate points in the large size limit, although this property has not been shown rigorously. Numerics for the WASEP are in agreement with this observation \cite{prolhac-these,prolhac-private}. This condensation property translated for the resolvent function $W_{L,N}(U)$ says that it converges to a limit $W(U)$ that has a finite number of cuts or poles in the complex plane, which correspond to the smooth curves or accumulation points formed by the $u_k$'s, and $W(U)$ is holomorphic everywhere else.

It leads us to the following assumption, on which all the computations of the next sections rely:
\begin{assumption}
\label{assumption1}
In the large $N,L$ limit with $N/L\to\rho_0$, $Ls\to\gamma$ and $L(p/q-1)\to 2\nu$, the resolvent $W_{L,N}(U)$ associated to the maximal eigenvalue converges to a well-defined limit $W(U)$ with a finite number of poles and cuts and holomorphic everywhere else. In the same limit, the function $\frac{R_{L,N}(U)e^{-sL/2}}{\sqrt{pq}^N(1+U)^L}$ converges to a well-defined limit $r(U)$ for all $U\neq -1$, that is holomorphic on $\mathbb{C}-\{-1\}$ and may have a singularity at $U=-1$ that can be an essential one.
\end{assumption}

If this is true, then \eqref{eq:bethepolyn:bis} has a limit given by:
\begin{equation}
\label{eq:bethe:resolvent}
r(U) = \cosh \left( \frac{\gamma}{2}+\rho\nu + \nu\frac{U}{1+U} -2\nu U W(U)\right) .
\end{equation}
In the weak asymmetry limit and under the previous assumption, the three equations \eqref{eq:translation:Qrep}, \eqref{eq:eigenvalueBethe:z} and \eqref{eq:R0} become
\begin{subequations}
\label{eq:propertiesW}
\begin{eqnarray}
 \gamma\rho + 2\rho\nu + 2\nu W(-1) \label{eq:normalisationW} &=& 0, \\ 
L\lambda_1(\gamma/L) \xrightarrow{L\to\infty} \tilde{\lambda}_1(\gamma) &=& 2\nu^2 \Big( W(-1)-W'(-1) \Big), \label{eq:eigenvalue:W} \\
r(0) &=& \cosh\left(\frac{\gamma}{2}+\rho\nu\right) \label{eq:r0cond}.
\end{eqnarray}
\end{subequations}
By construction, the resolvent $W(U)$ takes the generic form
\begin{equation}
\label{eq:decomp:W}
W(U) = \sum_{i=1}^{n_\text{poles}} \frac{\rho^{(i)}}{U-U^{(i)}} + \sum_{j=1}^{n_\text{cuts}} \int_{\Gamma_j} \frac{\rho^{(j)}(z)}{U-z} dz
\end{equation}
where $\rho^{(i)}>0$ is the fraction of roots condensed on the pole $U^{(i)}$ and $\rho^{(j)}(z)$ is the density of roots condensed at the point $z$ of the oriented cut $\Gamma_j$.
The resolvent $W(U)$ can be related to the Stieltjes transform of the spectral measure often introduced in random matrix theory, where the eigenvalues are replaced by the Bethe roots in the present case. One shall notice that the existence of poles for $W(U)$ is compatible with the assumption that the $u_k$'s are distinct for finite $L$ and $N$: finite size effects may split immediately the roots around the poles $U^{(i)}$. The density $\rho^{(j)}(z)$ and the fractions $\rho^{(i)}$ can be extracted from the knowledge of the resolvent $W(U)$ through:
\begin{eqnarray}
\rho^{(i)} &=& \lim_{U\to U^{(i)}} (U-U^{(i)}) W(U) \label{eq:polefraction:fromW}, \\
\rho^{(j)}(U)  &=&  \lim_{\epsilon\to 0^+, \text{$\epsilon$ along $\Gamma_j$}} \frac{1}{2i\pi}\Big( W(U-i\epsilon) - W(U+i\epsilon) \Big),\quad \forall U\in\Gamma_j . \label{eq:cutdensity:fromW}
\end{eqnarray}
In order to make the notations easier to read, it is useful to define for every $U$ on an oriented cut $\Gamma_k$ the two operators:
\begin{subequations}
\label{eq:Deltaop}
\begin{eqnarray}
\Delta^{(-)} f(U) &=& \lim_{\epsilon\to 0^+, \text{$\epsilon$ along $\Gamma_k$}} \Big( f(U-i\epsilon) - f(U+i\epsilon) \Big)\\
\Delta^{(+)} f(U) &=& \lim_{\epsilon\to 0^+, \text{$\epsilon$ along $\Gamma_k$}} \Big( f(U-i\epsilon) + f(U+i\epsilon) \Big)
\end{eqnarray}
\end{subequations}
and one has $\Delta^{(-)}f(U) = 2i\pi \rho^{(j)}(U)$ for $U\in\Gamma_j$.

\subsection{Characterization of the poles and the cuts}

As a limit of rational fractions with only one multiple pole at $U=-1$, the function $r(U)$ which appears in \eqref{eq:bethe:resolvent} is analytic on $\mathbb{C}-\{-1\}$. This implies that the r.h.s. of \eqref{eq:bethe:resolvent} can not have any cuts nor poles, excepted at $U=-1$. From the change of variable between the $z_k$'s and the $u_k$'s in the limit $L\to\infty$, it follows that the $u_k$'s can not be accumulated at $U=-1$ and thus $W(U)$ is well-defined at $U=-1$ (it can also be seen on \eqref{eq:normalisationW}). The only possible pole of $W$ compatible with the analyticity properties of $r(U)$ is at $U=0$. In the decomposition \eqref{eq:decomp:W}, at most one pole remains at $U=0$ with a residue $\rho_0\geq 0$.

Equation \eqref{eq:bethe:resolvent} can be rewritten as
\begin{equation}
\label{eq:def:Phi}
r(U)= \cosh \Phi(U)
\end{equation}
where $\Phi(U)$ contains the same amount of information as $W(U)$ in its cuts and poles. This function $\Phi(U)$ has a simple pole at $U=-1$ with residue $-\nu$ and has a finite limit for $U\to\infty$. Applying the operator $\Delta^{(-)}$ defined in \eqref{eq:Deltaop} to both sides of the previous equation gives for $U$ on a cut of $\Phi(U)$:
\begin{equation}
 0 = \sinh\left( \Delta^{(-)}\Phi(U)/2\right) \sinh\left( \Delta^{(+)}\Phi(U)/2\right). 
\end{equation}
This means that on any point $U$ of a cut of $\Phi$, one has either the condition
\begin{equation}
\label{eq:cond:minus}
\Delta^{(-)}\Phi(U) = 2 i\pi m, \quad m\in \mathbb{Z},
\end{equation}
or the second condition 
\begin{equation}
\label{eq:cond:plus}
\Delta^{(+)}\Phi(U) = 2i\pi n,\quad n\in\mathbb{Z}.
\end{equation}
The two functions $\Delta^{(\pm)}\Phi(U)$ are continuous along one cut, so the value of the integers $m$ or $n$ have to be constant along a cut, until one can jump from one condition to the other if there is a point where they are satisfied simultaneously. This remark is the key point of the identification of the phase transition in section \ref{sec:phasetransition}.

Conditions of the first type \eqref{eq:cond:minus} give directly through \eqref{eq:cutdensity:fromW} the density $\rho^{(j)}(z)$ on the cut of $W(U)$ where it is satisfied:
\begin{equation}
\rho^{(j)}(U) = \frac{m_j}{2\nu U}.
\end{equation}
The root density on the cuts for which \eqref{eq:cond:plus} is satisfied is more difficult to obtain; it can be performed by solving Cauchy principal value equations as in \cite{derrida-vanw-lecomte} but this tool will be useless in the simple case described in the next section.

Finding the set of cuts with the integers $m_j$ and $n_k$ corresponding to the largest eigenvalue $\lambda_1(s)$ is not a simple task. The next sections present the choice that gives the Gaussian part of the large deviation function and a candidate for the jammed phase in agreement with numerical simulations \cite{prolhac-these,prolhac-private}.

\section{Large deviation function of the current}

\subsection{Gaussian part of the current fluctuations}
\label{sec:gaussianpart}

It is known from the stochasticity of the matrix $\mathcal{W}^{(0)}$ that the rescaled eigenvalue $\tilde{\lambda}_1(\gamma)$ has to vanish when $\gamma\to 0$ and it corresponds to all the Bethe roots $z_j$ going to $1$, i.e. all the $u_j$'s going to $0$. The resolvent is then given by $W(U)=\rho/U$ when $\gamma=0$, which has a single pole at $U=0$ and no cuts.

As soon as $\gamma\neq 0$, cuts have to be opened around some points in the complex plane. Numerics \cite{prolhac-these,prolhac-private} tend to show that the largest eigenvalue $\lambda_1(\gamma)$ corresponds to the choice of a single cut $\Gamma$ for a discontinuity \eqref{eq:cond:plus} with $n=0$. The study of this cut is presented below and one gets easily convinced by similar computations that any opening of a second cut of any type \eqref{eq:cond:minus} or \eqref{eq:cond:plus} is impossible within the analyticity constraints on $\Phi$ and $r$, once assumption \ref{assumption1} is accepted. 

Let us introduce the extreme points $a_0$ and $a_1$ of the cut $\Gamma$, the polynomial $P_2(U)=\kappa(U-a_0)(U-a_1)$ such that $P_2(-1)=1$ and the function
\begin{equation}
\phi_2(U) = \sqrt{P_2(U)}
\end{equation}
which has the \emph{same} cut $\Gamma$ as $\Phi(U)$ going from $a_0$ and $a_1$ and the asymptotics $\phi_2(U)\sim U$ as $U\to\infty$ in order to define completely the sign of the square root. Then, if one assumes \eqref{eq:cond:plus} with $n=0$ on the only cut $\Gamma$, the product $\Phi(U)\phi_2(U)$ does not have any cut anymore on the complex plane by construction: the only remaining pole is at $U=-1$ and the asymptotic behaviour is $\Phi(U)\phi_2(U)\simeq O(U)$ at $+\infty$. One knows from complex analysis \cite{rudin} that it must take the form $S_2(U)/(U+1)$ where $S_2$ is a polynomial of degree at most $2$. Moreover, $r(U)$ has no other pole except at $-1$: $\Phi(U)$ cannot diverge for $U\to a_{0,1}$ and $P_2(U)$ has to divide $S_2(U)$. One gets finally:
\begin{equation}
\Phi(U) = \frac{\sqrt{P_2(U)}}{U+1}
\end{equation}
Only three parameters ($\kappa$, $a_0$ and $a_1$) are left. The properties \eqref{eq:propertiesW} of $W$ can be translated in terms of the function $\Phi(U)$:
\begin{subequations}
\label{eq:phi:expansions}
\begin{eqnarray}
\Phi(U) &\overset{U\to -1}{=}& -\frac{\nu}{U+1} + A_{-1} - \frac{\widetilde{\lambda}_1(\gamma)}{\nu} (U+1) + \mathcal{O}((U+1)^2) , \label{eq:expansion1}\\
\Phi(U) &\xrightarrow{U\to\infty} & A_\infty \label{eq:expansioninfty},\\
\cosh\Phi(0) &=& \cosh\left( \frac{\gamma}{2}+\nu\rho \right)  \label{eq:expansion0}
\end{eqnarray}
with the definitions of the two coefficients
\begin{eqnarray}
A_{-1} &=& \frac{\gamma(1-2\rho)}{2}+\nu(1-\rho) ,\\
A_\infty &=& \frac{\gamma}{2}+\nu(1-\rho) ,
\end{eqnarray}
\end{subequations}
and the values of the three unknown quantities $\kappa$, $a_0$ and $a_1$ can be computed easily. One obtains~: 
\begin{equation}
P_2(U) =  \nu^2 -2\nu A_{-1} (U+1) + A_\infty^2 (U+1)^2.
\end{equation} One also checks that \eqref{eq:expansion0} is simultaneously satisfied. Computing the eigenvalue $\widetilde{\lambda}_1(\gamma)$ is then an easy task, through an expansion of $\Phi(U)$ at the next order around $U=-1$:
\begin{equation}
\label{eq:gaussianlargedev}
\widetilde{\lambda}_1(\gamma) = \frac{A_\infty^2-A_{-1}^2}{2}  =  \frac{\rho(1-\rho)}{2} \Big(\gamma^2+2\gamma\nu\Big).
\end{equation}
It corresponds to the result expected from the macroscopic hydrodynamic theory \cite{bodineau-derrida}.

\begin{figure}
\includegraphics[width=11cm]{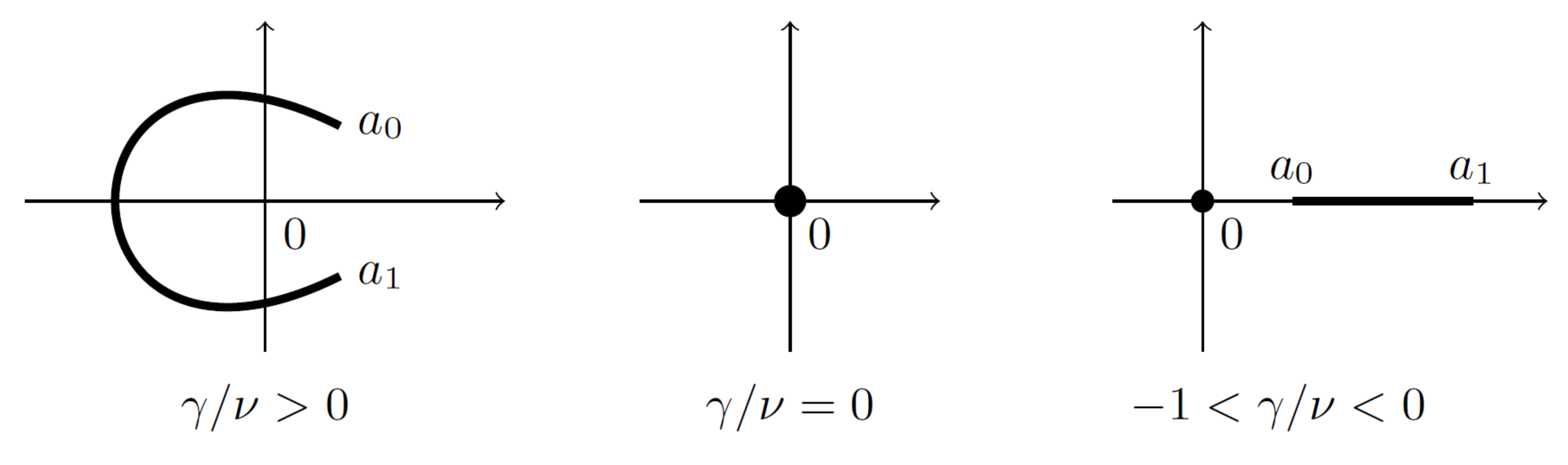}

\caption{\label{fig:typicalcuts}Condensation of the Bethe roots for different values of $\gamma$ in the Gaussian phase for $\rho=1/2$: for $\gamma/\nu>0$, the cut is symmetric under $z\to\bar{z}$ whereas it is real for $\gamma/\nu<0$. This second case $\gamma/\nu\leq 0$ also contains a pole at $0$. The densities along the cut in the Gaussian and the travelling wave phase are illustrated in figure \ref{fig:densitycuts}.}
\end{figure}

Although it has not been necessary for the computation of the eigenvalue, the shape of the cut of the Bethe roots can be extracted from $\Phi(U)$. The resolvent $W(U)$ is given by:
\begin{equation}
W(U) = \frac{1}{2\nu U}\left( \frac{\gamma}{2}+\rho\nu +\nu\frac{U}{U+1} - \Phi(U)\right).
\end{equation}
The point $U=0$ belongs to the cut if and only if $-\nu\leq \gamma \leq -2 \rho \nu$. For $\gamma\geq -2\rho\nu$, the residue at $0$ is given by~:
\begin{equation}
\label{eq:fracrootszero}
\rho_0=
\begin{cases}
0 & \text{if $\gamma/\nu>0$}, \\
\gamma/(2\nu) +\rho & \text{if $-2 \rho\nu \leq \gamma \leq 0$}.
\end{cases}
\end{equation}
The shape of the cut depends on the sign of $\gamma$, as shown on figure \ref{fig:typicalcuts}. For $\gamma>0$, the polynomial $P_2(U)$ has two complex conjugated zeros and the cut $\Gamma$ can be obtained by considering $\Delta^{(-)}W(U)=\rho_\Gamma(u)e^{i\theta(u)}$: the angle $\theta(u)$ gives the local direction of the cut and $\rho_\Gamma(u)$ gives the density of the roots. For $-2\rho\nu\leq \gamma<0$, the zeros and the cut are real and the density along the path is simply given by~:
\begin{equation}
\rho_\Gamma(u) =  \frac{\sqrt{ - P_2(u)}}{2u(u+1)}.
\end{equation}
For $-\nu\leq \gamma\leq -2\rho\nu$, the pole $U=0$ belongs to the cut and the density contains an additional Cauchy principal value (the pole at $U=0$ creates a local hole in the density, which produces this principal value). It is interesting to note that the eigenvalue, as well as the function $\Phi$, is not sensitive to the change of behaviour of $W(U)$ around $0$ but depends only on the polynomial $P_2(U)$ whose coefficients remain continuous as $\gamma$ varies.

The shape of the cut and the density of the roots along the cut gives a curious parallel with random matrices : the GUE ensemble of matrices with Gaussian entries has a spectrum given by the Wigner semi-circular law. Here, after change of variable, the Bethe roots have a semi-circular law whereas the fluctuations of the densities are expected (from the hydrodynamical limit) to be Gaussian. No understanding of this potential parallel is known however.

\subsection{The phase transition}
\label{sec:phasetransition}

As $\gamma$ varies, one checks that the creation of a small new cut of type \eqref{eq:cond:minus} or \eqref{eq:cond:plus} somewhere else in the complex plane is impossible within the analyticity constraints on $\Phi$ and $r$. However, there exists another scenario of modification of the cut obtained for the Gaussian part that gives a larger eigenvalue than \eqref{eq:hydro:gaussianlargedev} and \eqref{eq:gaussianlargedev} for $\gamma$ beyond a critical $\gamma_c$.

The previous cut was obtained by considering the case \eqref{eq:cond:plus}, which sets the value of $\Delta^{(+)}\Phi(U)$ to $0$ on the cut. The computation of $\Delta^{(-)}\Phi(U)$ on the same cut $\Gamma$ gives:
\begin{equation}
\Delta^{(-)} \Phi(U) = 2i\frac{\sqrt{-P_2(U)}}{U+1}.
\end{equation}
This quantity is equal to $2i\pi m$ for any $m\in\mathbb{Z}$ only on a discrete set of points on the cut. The trivial case $m=0$ corresponds to the edges of the cut where the density vanishes. The case $m=\pm 1$ is satisfied only if $\nu \sqrt{-P_2(U)} - \pi(U+1)$ vanishes on the cut, which correspond to a local density satisfying $\rho_\text{cut}(U)=1/(2\nu U)$. One checks that this is possible only if 
\begin{equation}
\label{eq:transitionpoint}
\frac{\rho(1-\rho)}{2} ( \gamma^2+2\gamma\nu) \leq -\frac{\pi^2}{2}.
\end{equation}
This threshold is precisely the one obtained from the linear stability analysis in the macroscopic hydrodynamic theory \cite{bodineau-derrida} but it is derived here only from complex analysis in a non-perturbative way.

\begin{figure}
\includegraphics[width=15cm]{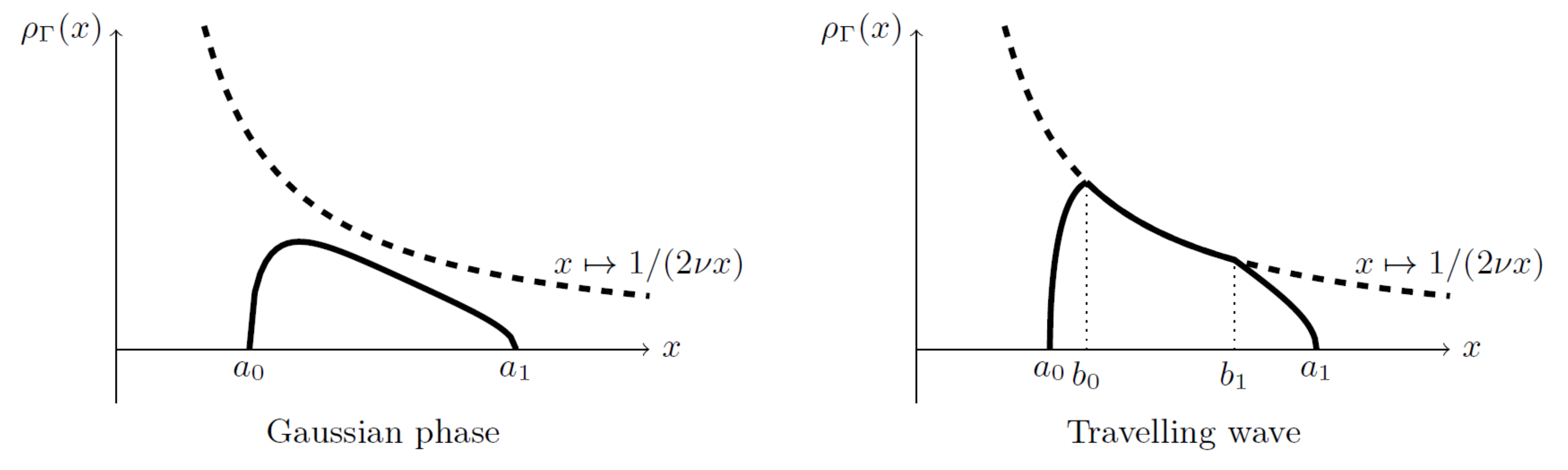}
\caption{\label{fig:densitycuts}Density of roots along the cuts in the Gaussian and the travelling wave phases (solid lines)~: compared to the first case (section \ref{sec:gaussianpart}), the cut in the second case (section \ref{sec:jammedphase}) is split in three parts and involves elliptic integrals.}
\end{figure}

At the transition point $\rho(1-\rho) ( \gamma^2+2\gamma\nu) = -\pi^2$, there exists a single point $b$ on the cut on which both constraints \eqref{eq:cond:minus} and \eqref{eq:cond:plus} are satisfied. For $\rho(1-\rho)( \gamma^2+2\gamma\nu) < -\pi^2$, the double point $b$ splits into two points $b_0$ and $b_1$, between which one can satisfy either \eqref{eq:cond:minus} or \eqref{eq:cond:plus} (see figure \ref{fig:densitycuts}). The second choice corresponds to the Gaussian case studied previously, whereas the first choice leads to another eigenvalue, which is computed in the next section.

The phase transition occurs necessarily when $\gamma\neq 0$ and all the cumulants of the current are not affected by the phase transition at the leading order in the system size. The phase transition can be seen in the cumulants only through the divergence of the finite-size correction, as described in \cite{derrida-vanw-lecomte,prolhac-cumul-weak}.

\subsection{Beyond the phase transition: a single hybrid cut}
\label{sec:jammedphase}

Beyond the transition, the candidate cut for $\Phi(U)$ is made of four singular points: the two extreme points $a_0$ and $a_1$ of the cuts and the two intermediate points $b_0$ and $b_1$ where the cut condition switches from \eqref{eq:cond:plus} to \eqref{eq:cond:minus}. The cut $\Gamma$ of $\Phi$ can be split into three parts~: $\Gamma_0$ (from $a_0$ to $b_0$) and $\Gamma_1$ (from $b_1$ to $a_1$), on which \eqref{eq:cond:plus} is satisfied with $n=0$, and 
$\Gamma_-$ between them (from $b_0$ to $b_1$), on which \eqref{eq:cond:minus} is satisfied with $m=1$ (see figure \ref{fig:densitycuts}).

As a generalization of the Gaussian case described previously, one can introduce the polynomial $P_4(U)= (U-a_0)(U-b_0)(U-b_1)(U-a_1)$ and the associate square root function $\phi_4(U)$ defined by
\begin{equation}
\phi_4(U) = \sqrt{P_4(U)}
\end{equation}
and whose first (resp. second) cut joins $a_0$ and $b_0$ (resp. $b_1$ and $a_1$) and coincides with the part $\Gamma_0$ (resp. $\Gamma_1$) of the cut of $\Phi(U)$ with type \eqref{eq:cond:plus}. The arbitrary sign of the square root is fixed by imposing $\phi_4(U)\sim U^2 $ as $U\to\infty$.

The derivative $\Phi'(U)$ satisfies the following cut conditions:
\begin{equation}
\begin{cases}
\Delta^{(+)}\Phi'(U) = 0 & \text{on $\Gamma_0$ and $\Gamma_1$},\\
\Delta^{(-)}\Phi'(U) = 0 & \text{on $\Gamma_-$}.
\end{cases}
\end{equation}
The function $\Phi'(U)\phi_4(U)$ is thus holomorphic on $\mathbb{C}-\{-1\}$, has a pole of order $2$ at $U=-1$ and is $O(1)$ as $U\to\infty$ by construction of $W(U)$: it has to be equal to $T_2(U)/(U+1)^2$ where $T_2$ is a polynomial of degree $2$. Around $-1$, we have the expansion $\Phi'(U)=\nu/(U+1)^2-\lambda/\nu+o(1)$ and thus $\Phi'(U)$ is given by:
\begin{equation}
\label{eq:derivativePhi}
\Phi'(U) = \nu \frac{\phi_4(-1)+(U+1)\phi'_4(-1)+\phi''_4(-1)(U+1)^2/2 -(\lambda/\nu^2)\phi_4(-1)(U+1)^2}{(U+1)^2\phi_4(U)}
\end{equation}
where $\lambda$ is the corresponding eigenvalue. The function $\Phi(U)$ can be written either as an integral of $\Phi'(U)$ or by the following formula obtained by analysing the cuts and poles of $\Phi(U)/\phi_4(U)$:
\begin{equation}
\label{eq:phi:reducedform}
\Phi(U) = \frac{\phi_4(U)}{U+1} \int_{b_0}^{b_1} \frac{z+1}{\phi_4(z)} \frac{1}{U-z}dz,
\end{equation}
which involves only the polynomial $P_4$ and satisfies directly both cut conditions \eqref{eq:cond:minus} and \eqref{eq:cond:plus}. Matching the previous expression \eqref{eq:phi:reducedform} with the required asymptotics \eqref{eq:phi:expansions} leads to:
\begin{subequations}
\label{eq:condP4:final}
\begin{eqnarray}
\label{eq:condAinfty}
\frac{\gamma}{2}+\nu(1-\rho) &=& \int_{b_0}^{b_1} \frac{z+1}{\phi_4(z)} dz , \\
\label{eq:condA1}
\nu &=& \int_{b_0}^{b_1} \frac{\phi_4(-1)}{\phi_4(z)} dz, \\
\frac{\gamma(1-2\rho)}{2}+\nu(1-\rho) &=&  -\int_{b_0}^{b_1} \frac{\phi_4(-1)+\phi_4'(-1)(z+1)}{(z+1)\phi_4(z)} dz, \label{eq:condA1bis} \\
\frac{\gamma}{2}+\nu\rho &=&  \int_{b_0}^{b_1} \frac{\phi_4(0)(z+1)}{z\phi_4(z)} dz, \label{eq:condA0} \\
 \frac{\tilde{\lambda}_1(\gamma)}{\nu}  &=& \int_{b_0}^{b_1} \frac{\phi_4(-1)+\phi_4'(-1)(z+1) + \phi_4''(-1) (z+1)^2/2}{(z+1)^2\phi_4(z)} dz  .\label{eq:P4lambda1}
\end{eqnarray} 
\end{subequations}
There are four independent constraints \eqref{eq:condP4:final} for four unknown quantities $a_0$, $b_0$, $b_1$, $a_1$ and the eigenvalue $\tilde{\lambda}_1(\gamma)$ is completely determined. The sign chosen for \eqref{eq:condA0} (left free by the condition \ref{eq:expansion0}) is set in order to be in agreement with the density of roots condensed in zero in the Gaussian phase \eqref{eq:fracrootszero}.

The expression of $W(U)$ can be recovered from \eqref{eq:phi:reducedform} and the definition of $\Phi(U)$. It gives the density of Bethe roots along the cut:
\begin{equation}
\rho_\Gamma(u) = \begin{cases}
\displaystyle{\frac{1}{2\nu u}}& \text{for $u\in [b_0,b_1]$}, \\
\\
\displaystyle{\frac{\sqrt{|P_4(u)|}}{4\pi\nu u (u+1)} \int_{b_0}^{b_1} \frac{1}{\sqrt{P_4(z)}} \frac{z+1}{|z-u|}dz} & \text{for $u\in[a_0,b_0]\cup[b_1,a_1]$},
\end{cases}
\end{equation}
and the density of roots $\rho_0$ condensed at $u=0$:
\begin{equation}
\rho_0 = \frac{\gamma+2\rho\nu}{4\nu} - \frac{1}{2\nu} \int_{b_0}^{b_1} \frac{\sqrt{|P_4(0)|}}{\sqrt{P_4(z)}}\frac{z+1}{z} dz  .
\end{equation}
The case corresponding to a cut including $0$ and a negative fraction $\rho_0$ can be treated as in the Gaussian case by considering a Cauchy principal value of the quantities written above; it does not lead to any additional phase transition nor any relevant non-analyticity.

\begin{figure}
\begin{center}
 \input{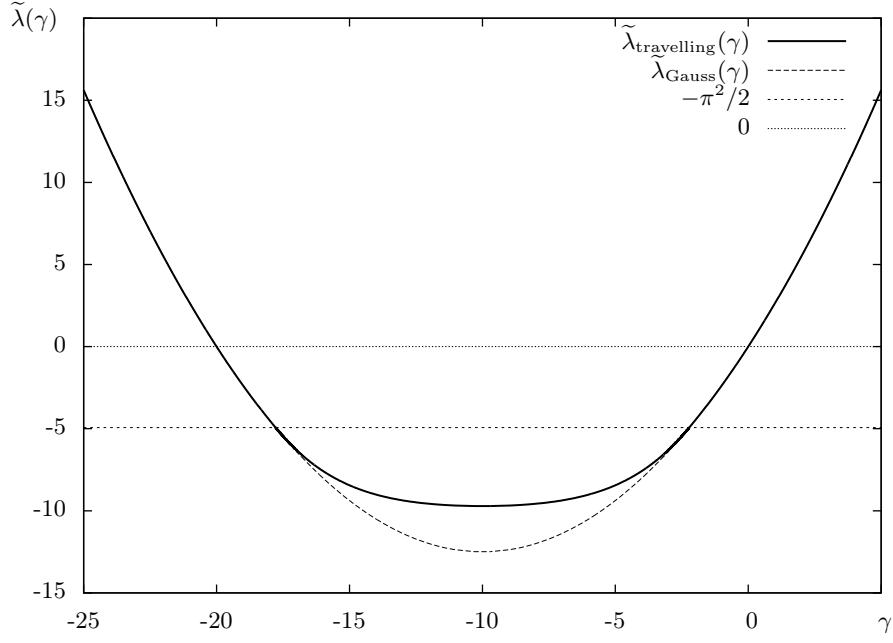}
\end{center}
 \caption{\label{fig:plotslargedev}Legendre transform of the large deviation function for $\rho=1/2$ and $\nu=10$. This plot shows that the hybrid cut with both $\Delta^{(+)}$ and $\Delta^{(-)}$ conditions gives a larger eigenvalue as soon as $\widetilde{\lambda}(\gamma)<-\pi^2/2$.} 
\end{figure}

\subsection{Partial correspondence with the hydrodynamic description}
\label{sec:relationtravelling}

Numerics show that the expression of the eigenvalue we obtained coincides with the one obtained from the hydrodynamical approach described in appendix \ref{app:profiles}. However, the exact mapping between the polynomial $P_4$, which satisfies the set of equations \eqref{eq:condP4:final}, and the polynomial $R_4$, which satisfies the set \eqref{eq:hydroeqs}, seems to be non-trivial (at least more complicated than just an homographic transformation).

For the particular case $\rho=1/2$ however, the mapping is exact and simple. For a half-filled system, the speed $v$ of the travelling wave in appendix \ref{app:profiles} is $v=0$ and the profile is symmetric under $g \to 1-g$ (particle-hole duality): we obtain the values $C_2=0$ and $g_1=1-g_2$ (with the notation of appendix \ref{app:profiles}). This particularity emerges as well in the Bethe Ansatz approach and corresponds to an additional symmetry of the polynomial $P_4(U)$. 

In the Gaussian case, the polynomial $P_2$ is given by 
\begin{equation}
 P_2(U) = -\nu^2 U + \left(\frac{\gamma+\nu}{2}\right)^2 (U+1)^2
\end{equation}
and is invariant under the additional symmetry $P_2(U) \to U^2 P_2(1/U)$. 

In the non-Gaussian case, the polynomial $P_4(U)$ shares the same additional symmetry $P_4(U) \to U^4 P_4(1/U) = P_4(U)$. The equation for $A_{-1}$ fixes one of the coefficients  such that $P_4(U)$ can be written as:
\begin{equation}
 P_4(U) = p U^2 + p' U(U+1)^2 +  (U+1)^4 = (U+1)^4 \tilde{P}\left( \frac{U}{(U+1)^2} \right).
\end{equation}
The two coefficients $p$ and $p'$ satisfy
\begin{align}
 \frac{\gamma+\nu}{2} &= \int_{b_0}^{1/b_0} \frac{u+1}{\sqrt{P_4(u)}} du, \\
 \nu &= \int_{b_0}^{1/b_0}\frac{\sqrt{P_4(-1)}}{\sqrt{ P_4(u)}} du,
\end{align}
which are equivalent to \eqref{app:eq:Jv1} with $v=0$ and \eqref{app:eq:totalmass} after the change of variable $g=U/(U+1)$ and a different normalization of $P_4$; the eigenvalue is the same in both cases.

The full correspondence between both approaches for $\rho \neq 1/2$ is however not clear and requires further investigation, as well on the correspondence between the polynomials $P_4$ and $R_4$ as on the relation between the real travelling wave profile and the curves of the Bethe root.

A second interesting explicit correspondence corresponds to the limit $\nu\to\infty$ with the scaling $\gamma=\nu\bar{\gamma}$; this limit is expected to coincide with the TASEP large deviation function as already explained in \cite{bodineau-derrida}. The divergences that appear in equations \eqref{eq:condP4:final} imply that $a_0 \uparrow b_0$ and $a_1 \downarrow b_1$ in this scalings. In terms of the characteristics of the cut, it means that the two parts with $\Delta^{(+)}\Phi=0$ shrink to zero and only remains the cut with $\Delta^{(-)}\Phi=2i\pi$. Two methods lead to the same result: either one extracts carefully the asymptotics of the all right terms \eqref{eq:condP4:final} or one constructs directly $\Phi$ with the prescribed limit behaviour. One obtains for any $U$ not to close to the cut
\begin{equation}
 \Phi(U) \sim \nu \left[ \alpha_0+\frac{\alpha_{-1}}{U+1} + \ln\left(\frac{U-b_1}{U-b_0}\right) \right] + o(\nu)
\end{equation}
Satisfying the renormalized conditions \eqref{eq:phi:expansions} leads to $\alpha_{-1}=0$, $\alpha_{0}=\bar{\gamma}{2}+(1-\rho)$ and
\begin{equation*}
 \ln\left( \frac{1+b_1}{1+b_0}\right)= -\bar{\gamma}\rho , \quad \ln\left(\frac{b_1}{b_0}\right)= -\bar{\gamma}
\end{equation*}
It leads to the scaling of the eigenvalue:
\begin{equation}
 \lim_{\nu\to\infty,\gamma/\nu\to\bar{\gamma}} \frac{\widetilde{\lambda}_1(\gamma)}{\nu^2} = - \frac{(1-e^{\bar{\gamma}\rho})(1-e^{\bar{\gamma}(1-\rho)})}{1-e^{\bar{\gamma}}}
\end{equation}
for all $\bar{\gamma}<0$, and one recovers the asymptotics of the TASEP as explained in \cite{bodineau-derrida}. It is interesting to see that the coexistence of high and low density domains in the TASEP behaviour corresponds to $\Delta^{(-)}$ cuts, whereas the gaussian fluctuations around a flat profile correspond to $\Delta^{(+)}$ cuts and it would be of interest to understand this correspondence from the eigenvector \eqref{eq:betheansatz}  itself.

\subsection{Existence of other phase transitions : a short discussion}

A question left open in \cite{bodineau-derrida} is the existence and the nature of other phase transitions in the WASEP that may overcome the phase transition described in the previous sections.

The standard scenario in Bethe Ansatz (as well as in random matrices) is a condensation of the Bethe roots on curves or isolated points in the thermodynamic limit $N,L\to\infty$ with $N/L\to\rho$. These condensation curves correspond to cuts of the resolvent $W$. Phase transitions occur when the number or the topology of the curves vary: splitting of a curve into two, emergence of a second curve, closure of a cut, etc.

The scenario presented here is different: no change occurs in the number of cuts nor in their topology (and the pole of $W(U)$ at $0$ is irrelevant) but a phase transition still occurs due to the change of type of the Riemann surface associated to the function $\Phi(U)$. Such a phenomenon does not appear in random matrices (GUE with a potential for example) since the resolvent of the eigenvalues satisfies a quadratic equation much simpler than the $\cosh$ equation \eqref{eq:def:Phi}. The Riemann surface of the $\text{argcosh}$ functions gives further possibilities, which are well described by the set of conditions \eqref{eq:cond:minus} and \eqref{eq:cond:plus}. 

As already announced in section \ref{sec:gaussianpart}, it is not possible to open other cuts of type \eqref{eq:cond:plus}. One remarks first that the function $\Phi'(U)$ has cuts only of type \eqref{eq:cond:plus}. If we suppose the existence of $n$ cuts with end points $a_j$ and $a'_j$ and construct the polynomial $p_{2n}(U)$ whose zeros are these end points, then $f(U)=\Phi'(U)\sqrt{p_{2n}(U)}$ has no cut if one makes the cuts of the square root coincide with the ones of $\Phi'(U)$. Thus, $f$ is analytical on $\mathbb{C}$ except at $U=-1$ where it has a pole of order $2$. Moreover, $f(U)=O(U^{-2})$ for $U\to\infty$. The only choice corresponds to $f(U)=R_n(U)/(U+1)^2$ where $R_n(U)$ is a polynomial of degree $n$. The analyticity properties of $\Phi$ set then $R_n(U)=0$ except for $n=1$, which corresponds to the Gaussian case. Opening a cut with \eqref{eq:cond:minus} only is also impossible. Thus, there appears to be no phase transition due to a change in the number of cuts.

However, within the one-cut assumption, one could imagine a hybrid cut as in the previous section but with the condition \eqref{eq:cond:minus} for $|m|\geq 2$ or even a hybrid cut made of successive paths alternating between the two conditions \eqref{eq:cond:minus} and \eqref{eq:cond:plus}, generalizing the scheme presented in the previous section. Checking if it is possible or not requires a precise study of the integrals of the type \eqref{eq:condP4:final} that would emerge with a polynomial of larger degree: it is beyond the target of the present paper but it would be very interesting to check it rigorously.

The discussion of the existence of several curves presented in this section is valid only for part of the eigenvalues  (more precisely, the ones for which the assumption \ref{assumption1} is valid). When the asymmetry $p-q$ is reduced from the PASEP to the WASEP, the largest eigenvalue as well as some others follows the scaling \eqref{eq:hydro:gaussianlargedev} for which assumption \ref{assumption1} holds for the Bethe roots. The other part of the spectrum does not have to satisfy this assumption and this scaling form: with another well-adapted scaling, the Bethe roots may still condensate on some curves in arbitrary number.

\section{Conclusion}

The exclusion process in the weak asymmetry regime $p-q=O(1/L)$ shows a phase transition in the large deviation function of the current that was first predicted in \cite{bodineau-derrida}: this paper presents an alternative analysis of this transition with the Bethe Ansatz toolbox.

Different types of results have been obtained in the present paper and each of them leads to some new questions. From the point of view of integrable systems, the first interesting point is the shape of the condensation curve of the Bethe root in the thermodynamic limit considered in this paper: to our knowledge, it is the first example of a single cut with a piecewise-defined continuous density. Its origin is the non-trivial $\cosh$ equation satisfied by the resolvent instead of the standard quadratic equations.

One of the interest of the scaling $p-q=O(1/L)$, which is unusual for spin chains, is the existence of an hydrodynamic limit that leads to partial differential equations that are classically integrable whereas the microscopic Markov transition matrix satisfies quantum integrability (see section \ref{sec:hydro}). The present example gives another example of transition from quantum to classical integrability.

The main progress that leads to the solution presented here is the formulation (\ref{eq:bethepolyn:bis},\ref{eq:betheeqs:newdefs}) of the Bethe equations that is suitable to the scaling limit $p-q=O(1/L)$. Variations on this procedure should also lead to interesting results \cite{dsimon-inprogress} for other weak asymmetry regimes such as the KPZ scaling regime $p-q=O(1/\sqrt{L})$, which is studied actively for different geometries \cite{bertini-giacomin,prolhac-these,sasamoto-spohn,quastel}. Higher order perturbative expansions should give access to the finite size corrections of the large deviation function. It would be particularly interesting to characterize more precisely the phase transition (exponents, scaling of the finite size corrections, etc.).

Some open questions also emerge from the present paper. Although the same integrals that determine the large deviation function of the current appear in both the Bethe Ansatz and the hydrodynamic approaches, the precise correspondence between the contour of the Bethe roots and the density profile of the travelling wave is still missing: it would be interesting to understand the relation between them and to establish a "dictionary" between corresponding notions in both approaches. A corollary would be the correspondence between the two fourth-degree polynomial $P_4$ and $R_4$ that emerge in the two approaches.

The last new information that comes with the Bethe Ansatz is the knowledge of the eigenvectors, which also describe in principle the spatial correlations. Computing correlation functions from the Bethe Ansatz is known to be a difficult problem in general; nevertheless, the macroscopic hydrodynamic description tends to show that the macroscopic correlations should have a simple description. It is probable that the coordinate Bethe Ansatz form of the eigenvectors should have a simpler limit than usual in the present weak asymmetry regime but it has not been possible to study it rigorously yet.

\section*{Acknowledgments}

I am deeply grateful to Bernard Derrida, who introduced me to this problem in 2005, for all the old and the recent discussions we shared on this subject. I am also grateful to Sylvain Prolhac for showing me some of his unpublished numerical results about the density of the Bethe roots, and to Thierry Bodineau for all our discussions.

\appendix

\section{Optimal travelling wave profile of the variational principle}
\label{app:profiles}

The normalized Legendre transform \eqref{eq:legendre} of the large deviation function of the current is obtained from the variational principle \eqref{eq:actionI}:
\begin{equation}
\label{eq:legendreoptim}
\widetilde{\lambda}_1(\gamma) = \sup_{J,\rho(x,\tau)} \left( \gamma J - \lim_{T\to\infty} \frac{\mathcal{I}_{[0,T]}(j,\rho)}{T} \right)
\end{equation}
where the fluctuating profile $\rho$ and current $j$ satisfy the conditions \eqref{eq:currentfixed} and \eqref{eq:massconserv} with $j_0=J$. The optimal travelling wave studied in \cite{bodineau-derrida} corresponds to the choice of a particular type of time-dependent profiles
\begin{equation}
\rho(x,t) = g(x-vt).
\end{equation}
One obtains $j(x,t)=J+v (g(x-vt)-\rho)$ from \eqref{eq:currentfixed} and \eqref{eq:massconserv}, and \eqref{eq:legendreoptim} becomes time-independent:
\begin{equation}
\label{eq:legendrewave}
\widetilde{\lambda}_1(\gamma) = \sup_{J,g,v} \left( \gamma J - 
\int_0^1 \left[ \frac{(J+v(g(x)-\rho) -\nu \sigma(g(x)))^2}{2\sigma(g(x))}+\frac{g'(x)^2}{8\sigma(g(x))} \right]dx \right).
\end{equation}
The previous functional is invariant by translation of the profile $g$ and its optimization yields (see \cite{bodineau-derrida}):
\begin{equation}
g'(x)^2 = 4(J+v(g(x)-\rho) -\nu \sigma(g(x)))^2 - 4(C_1+C_2 g(x))  \sigma(g(x)) =  4 R_4^{(J,v)}(g(x))
\end{equation}
where $C_1$ and $C_2$ are two constants that are fixed once the extremal densities $g_1$ and $g_2$ of $g(x,t)$ over $[0,1]$ are fixed by the two conditions:
\begin{subequations}
\label{eq:hydroeqs}
\begin{eqnarray}
\label{app:eq:totalmass}
1 &=& \int_0^{1} dx = \int_{g_1}^{g_2} \frac{1}{\sqrt{R_4^{(J,v)}(g)}} dg, \\
\rho &=& \int_0^{1} g(x) dx = \int_{g_1}^{g_2} \frac{g}{\sqrt{R_4^{(J,v)}(g)}} dg.
\end{eqnarray}
The change of variable $x\mapsto g(x)$ for the two intervals over which $g$ is monotonic yields:
\begin{equation}
\widetilde{\lambda}_1(\gamma) = \sup_{J,v} \left( \gamma J - \frac{1}{2} 
(C_1+C_2\rho) -\int_{g_1}^{g_2} \frac{1}{\sigma(g)} \sqrt{R_4^{(J,v)}(g)}\right) .\\
\end{equation}
The optimal values of $J$ and $v$ are then obtained implicitly by:
\begin{eqnarray}
\label{app:eq:Jv1}
\gamma+\nu = J \int_{g_1}^{g_2}\frac{1}{\sigma(g)\sqrt{R_4^{(J,v)}(g)}}dg + v \int_{g_1}^{g_2} \frac{g-\rho}{\sigma(g)\sqrt{R_4^{(J,v)}(g)}}dg ,\\
\label{app:eq:Jv2}
0 = J \int_{g_1}^{g_2} \frac{g-\rho}{\sigma(g)\sqrt{R_4^{(J,v)}(g)}}dg + v \int_{g_1}^{g_2} \frac{(g-\rho)^2}{\sigma(g)\sqrt{R_4^{(J,v)}(g)}}dg .
\end{eqnarray}
\end{subequations}

For $\rho=1/2$,  the speed is $v=0$ and the correspondence with the Bethe Ansatz calculation presented in section \ref{sec:relationtravelling} is given by $J=\nu\sqrt{p_4}$ and $C_1-2\nu J=p_3$.

\end{document}